\documentclass[draft]{agujournal2019}
\usepackage{url} 
\usepackage[inline]{trackchanges} 
\usepackage{soul}
\draftfalse

\journalname{Earth's Future}

\begin{document}

%
%


\title{Tradeoffs inherent in solar geoengineering peak-shaving strategies}

%
%




\authors{Thomas Hornigold\affil{1}, Myles Allen\affil{2}}


\affiliation{1}{Department of Atmospheric, Oceanic and Planetary Physics, University of Oxford}
\affiliation{2}{Environmental Change Institute, University of Oxford}





\correspondingauthor{Thomas Hornigold}{thomas.hornigold@linacre.ox.ac.uk}




\begin{keypoints}
\item Solar geoengineering ``peak-shaving" strategies are considered in the abstract.
\item We find robust trade-offs between the suppressed temperature overshoot and the required duration of SRM across a range of idealised and modelled scenarios.
\item The cost of, and committment to, long-term and large-scale negative emissions must be considered in assessing the cost of any finite-duration geoengineering scheme. 
\end{keypoints}

%
%

%
%


\begin{abstract}

Some have suggested solar geoengineering, cooling the planet by reflecting sunlight, could be deployed temporarily to ``shave the peak" of global temperatures. This would use limited deployment to keep global mean temperatures below a given threshold, buying time for emissions reductions and negative emissions to reduce greenhouse gas concentrations. However, we suggest the trade-offs inherent in such scenarios are not widely appreciated. In this study, we explore general features of such scenarios, using a combination of simple analytical relationships between emissions and temperatures and modelling to illustrate the strong trade-off between the duration of geoengineering deployment and the temperature overshoot that is suppressed by geoengineering, as well as the scale of negative emissions that are required. Our results indicate that, unless climate sensitivity is at the high end of estimates, shaving a substantial peak – say, of 2K to 1.5K – would require at least a century of deployment of solar geoengineering and large-scale negative emissions, under a broad range of emissions trajectories. If negative emissions are not available at the scale of tens of gigatonnes of CO\textsubscript{2} per year, substantial peak-shaving will require deployment of geoengineering for multiple centuries. Our analysis emphasise the importance of factoring in the scale, feasibility, and cost of the negative emissions required to avoid indefinite commitment to solar geoengineering.

\end{abstract}

\section*{Plain Language Summary}

Scientists are confident that human emissions of greenhouse gases are warming the planet. Some have suggested that solar geoengineering, where particles are put into the atmosphere to reflect sunlight and cool the planet, could reduce the negative effects of this climate change. Small amounts of geoengineering may have more benefits and fewer risky side-effects than larger amounts. For example, we could use geoengineering to keep to below a temperature threshold, like those agreed upon in the Paris Climate Accord. In this paper, we use mathematical equations and computer models to explore a range of ways that this could be done. We show that there is a trade-off. In cases where you only need geoengineering for a short time, this is unlikely to have much of an effect on temperatures. Or, you can reduce temperatures by a larger amount, but then you must keep geoengineering going for many decades, which could be difficult politically. We also note that, because greenhouse gases remains in the atmosphere for centuries, if you want to stop the geoengineering, you must remove billions of tonnes of carbon dioxide. We argue that these trade-offs should be carefully considered and made apparent in discussions of geoengineering.

%
%

%


%
%
%
%

\section{Introduction}

Under what circumstances could we consider solar geoengineering? Many authors have observed that solar geoengineering cannot be a substitute for mitigation \cite{goes2011economics} \cite{keith2016solar}.  Solar radiation management (SRM) imperfectly compensates for the effects of greenhouse gases, leading to residual climatic changes. These residual climate changes are likely to be worse as the amount of radiative forcing from SRM is increased. For example, SRM reduces precipitation more quickly than temperature. This is because the net result of cooling the surface with SRM, while the troposphere is warmed by greenhouse gases, is an atmosphere that is more stable to convection. Consequently, trade-offs exist between regions and between returning temperature and precipitation to preindustrial baselines using SRM. \cite{moreno2012simple} SRM would do little to address ocean acidification \cite{williamson2012ocean}, and may result in residual polar ocean warming or changes to large-scale circulation \cite{fasullo2018persistent}. The greater the radiative forcing that is offset by SRM, the more dangerous it would be to suddenly cease its use due to the termination effect: a rapid rebound of temperatures \cite{jones2013impact}. For this reason, we suggest that the temperature change compensated for by geoengineering is referred to as ``suppressed" rather than ``avoided". In addition, the magnitude and latitudinal distribution of the radiative forcing that can be obtained with the most viable method, stratospheric aerosol injection, is not yet clear, with model studies in disagreement \cite{niemeier2015limit} \cite{kleinschmitt2018sensitivity}. A scenario where humanity is committed to ever-increasing deployment of SRM while emissions continue to climb is evidently fraught with risk.

“Peak-shaving” scenarios, where SRM is used temporarily to keep temperatures below a threshold – for example, the 1.5C or 2C targets of the Paris Agreement – have been suggested as a more reasonable form of deployment, such as in the IPCC SR1.5 report (Section 4.3.8 and Cross-Chapter Box 10) \cite{de2018strengthening} \cite{macmartin2018solar}, which would buy time for mitigation to take effect. Modelling suggests that limited deployment of SRM - for example, to cancel half of the warming due to greenhouse gases - could reduce the negative residual climatic changes in most regions. \cite{irvine2019halving} Here, we do not seek to evaluate whether the climate benefits of such peak-shaving outweigh the risks, but instead to explore the parameter space of peak-shaving scenarios.

\medskip


First, we note that if SRM is to be deployed temporarily, CO\textsubscript{2} emissions must peak and decline to net zero. This follows from the approximate proportionality of temperature changes to cumulative carbon emissions \cite{allen2009warming}. Once emissions reach net zero, carbon dioxide concentrations in the atmosphere will gradually decline due to ocean uptake and weathering, but substantial net-negative CO\textsubscript{2} emissions are likely to be necessary to avoid centuries of commitment to SRM in a peak-shaving scenario.

\section{Peak-shaving in the abstract}
\subsection{A toy example: parabolic peak}

As a simple example, consider only CO\textsubscript{2} emissions. Assume that no mitigation occurs until a particular threshold temperature - say, 1.5C - is crossed. Thereafter, CO\textsubscript{2} emissions begin falling at a constant rate, becoming net-negative emissions at this same rate. Under these circumstances, the temperature peak will be approximately parabolic, as the emissions in a given year are approximately proportional to the rate of temperature change in that year. Let D be the duration of peak-shaving above a given temperature threshold, T be the temperature excess above threshold, and R be the rate of warming when the threshold is first exceeded (noting that this is approximately the Transient Climate Response to Cumulative Emissions, TCRE, multiplied by the emissions when the threshold is exceeded.) 

\medskip

We can express this quadratic curve as $ T(t) = \alpha t (t – D) $, where D is the length of time for which the threshold is exceeded. Differentiating gives $ dT/dt = \alpha (2t – D)$, and $R = - \alpha D$ by definition. The maximum temperature above the threshold, or the total overshoot that would be suppressed by SRM, which we denote $T_{so}$, is equal to $– \alpha D^2 / 4.$ We can then express the suppressed temperature overshoot in terms of the duration and the warming rate when the threshold is first crossed: $T_{so} = RD/4$ or, equivalently, $4T_{so}/R = D.$

\medskip

In such an idealised scenario, the duration of the SRM commitment is twice the time taken to reach net zero. The attainable mitigation rate therefore determines the cumulative negative emissions required, and the duration of commitment to SRM. When this is combined with the climate sensitivity - for example, through the rate of warming - the temperature overshoot can be inferred. 

\medskip

This allows us to see the tradeoff between the implied duration of deployment and the suppressed temperature overshoot. For example, at the current rate of warming of 0.25K per decade, shaving a peak of 2C to 1.5C implies a commitment to deployment of around 80 years. (In reality, this is likely to be an underestimate, as we will explain.) 

\medskip

Considered in the context of the Paris Agreement, current scenarios that limit global warming to 1.5K (or 2K) include overshoot scenarios where temperature temporarily rises above the target, before falling at the end of the century. SRM could be deployed to suppress such an overshoot, as in SR1.5 Section 4.3.8. In the example considered in the IPCC report, overshoot of around 0.1K is avoided.  Assuming rates of warming and cooling are similar to today’s magnitude of warming at 0.25K/decade, the rule-of-thumb implies a commitment of 32 years, which is similar to the indicative time-scale shown (35 years). This exemplifies the trade-off between the duration of the implied commitment to SRM and the magnitude of the suppressed temperature overshoot.

\medskip

By integrating, we can determine the area under the curve and hence the number of degree–years avoided, as a rough metric for avoided damages \cite{smith2013long}, although due to SRM's imperfect compensation for increased GHG concentrations, this should not be considered fungible with the equivalent temperature increase avoided by mitigation. This gives $RD^{2}/6$ and consequently would equate to around 267 degree–years avoided in the above example. 

\medskip

Even this simple example illustrates the trade-off between duration of the commitment and the suppressed temperature overshoot for a given rate of warming. The more warming we use SRM to avoid – and hence, the more damaging any potential termination – the longer our commitment must be.

\subsection{Different rates of mitigation and net-negative emissions deployment}

This analysis can be simply extended to consider the more general case for overshoot trajectories where the rate of warming when the target is first crossed and the rate of cooling are not the same, but parabolic trajectories are assumed either side of the peak - in other words, mitigation and net-negative emissions deployment occur at different rates. 

In this case, we have (where warming and cooling rates are positive-definite magnitudes):

\medskip

\begin{equation}
D_{warm} = \frac{2T_{so}}{R_{warm}} 
\end{equation}

\begin{equation}
D_{cool} =  \frac{2T_{so}}{R_{cool}}
\end{equation}

\begin{equation}
D_{overshoot} = \frac{2T_{so}}{R_{warm}}  \left( 1 + \frac{R_{warm}}{R_{cool}} \right) 
\end{equation}

\medskip

If the rate of warming on the way up exceeds the rate of cooling on the way down, then overshoot is longer. Given that rates of warming and cooling are approximately proportional to the emissions of CO\textsubscript{2}, the ratio of $ R_{warm} $ to $ R_{cool} $ is roughly the ratio of emissions for the year when the carbon budget is exceeded and the negative emissions when the ``carbon debt” has been repaid. 

\medskip

The generalised formula for degree-years is similarly obtained:

\begin{equation}
D Y_{overshoot}=\frac{4 T_{so}^{2}}{3 R_{warm}}\left(1+\frac{R_{warm}}{R_{cool}}\right)
\end{equation}

If we denote the carbon budget overshoot by $ \Delta C_{ex} $, and emissions as a function of time as $ E\left(t\right) $ then we have, with $ \lambda $ as the TCRE: 

\medskip

\begin{equation}
D \approx 2 \cdot \lambda \cdot \Delta C_{e x}\left(\frac{1}{\lambda  \cdot E\left(t_{warm}\right)}+\frac{1}{\lambda  \cdot\left|E\left(t_{cool}\right)\right|}\right)
\end{equation}

\begin{equation}
D \approx  2 \cdot \Delta C_{e x}\left(\frac{1}{E\left(t_{warm}\right)}+\frac{1}{\left|E\left(t_{cool}\right)\right|}\right)\end{equation}

\medskip

We see from this that the duration of the overshoot – and hence the implied SRM commitment – depends on the amount the carbon budget is exceeded, the timescale over which you continue to emit carbon dioxide and warm, and the timescale to remove the full anthropogenic excursion of CO\textsubscript{2}. In this approximation, the dependence on climate sensitivity arises from its implications for the carbon budget for 1.5K or 2K. Higher climate sensitivity reduces the cumulative emissions required for a given overshoot, which shortens both the timescale of continued emissions and the timescale of negative emissions required to remove them.

\subsection{The implications of a negative emissions floor}

Suppose that global net-negative emissions can only be increased up to a certain ``floor" (n). This could be due to geobiophysical limitations in the case of large-scale BECCS, afforestation, or enhanced weathering, or societal willingness to pay in the case of Direct Air Capture. We might also anticipate that the average rate at which large-scale net-negative emissions are deployed (m\textsubscript{2}) will differ from the average rate of mitigation (m\textsubscript{1}). If mitigation and geoengineering start when the temperature threshold is crossed, and carbon emissions fall linearly from a maximum of C\textsubscript{max},  then our carbon budget formulation gives us:

\begin{equation}
\Delta C_{ex} = \frac{1}{2} \frac{C^{2}_{max}}{m\textsubscript{1}}   
\end{equation}

With t\textsubscript{neg} as the total time committment to any negative emissions, and t\textsubscript{floor} the time commitment to negative emissions at their lowest value, we have:

\begin{equation}
 t_{neg} = \frac{n}{m\textsubscript{2}} + t_{floor} 
\end{equation}

Requiring that the total negative emissions cancel out the exceedance of the carbon budget, we have:

\begin{equation}
 \Delta C_{ex} = \frac{1}{2} \frac{n^{2}}
{m\textsubscript{2}} + t_{floor} n   
\end{equation}

Solving for t\textsubscript{floor} and with the total duration of the overshoot as D, we obtain:

\begin{equation}
 D = \frac{C^{2}_{max}}{2m\textsubscript{1} n} + \frac{C_{max}}{m\textsubscript{1}} + \frac{n}{2m\textsubscript{2}}
\end{equation}

Which simplifies in the special case that m\textsubscript{1} = m\textsubscript{2} = m to:

\begin{equation}
D = \frac{1}{2nm} (C_{max} + n)^2 
\end{equation}

We also have that

\begin{equation}
 T_{so} = \Delta C_{ex} \cdot \lambda  = \frac{C^{2}_{max}\cdot \lambda}{2  \cdot m\textsubscript{1}} 
\end{equation}

\medskip

Which allows us to re-express the duration in terms of the TCRE, the peak shaved, the negative emissions floor, and the mitigation rate, if desired.

\subsection{The double exponential emissions trajectory}

An alternative model of emissions trajectories uses a double exponential function. For peak-shaving emissions trajectories, subject to  a negative emissions floor, you can model the emissions after mitigation begins as:

\begin{equation}
E(t) =  C_{max} \exp(-  m\textsubscript{1} t) - n ( 1 - \exp( - m\textsubscript{2} t)) 
\end{equation}

where

\begin{equation}
m\textsubscript{1,2} = \ln(2)/t\textsubscript{1,2} 
\end{equation}

and t\textsubscript{1,2} are the timescales for halving (gross) emissions and scaling up the deployment of negative emissions to half of its maximum value, respectively. Alternatively, to represent emissions which change by x\% per year, one could consider m\textsubscript{i} = ln(1+0.01x). In this idealised emissions scenario, rapid mitigation is combined with expanding deployment of negative emissions at scale to reach net zero, while both mitigation rate and negative emissions deployment rate slow down over time. This corresponds to idealised trajectories where emissions fall by a given average percentage each year. 

\medskip

This models well the trajectory of the Low Energy Demand scenario \cite{grubler2018low} and the rapid decarbonisation pathway referred to as the ``Carbon Law'', which proposes halving emissions each decade while scaling up negative emissions to a gigatonne scale. \cite{Rockstrom1269}

\medskip

This gives a cumulative exceedance of the carbon budget as:

\begin{equation} \Delta C_{ex} = \frac{C_{max}}{m\textsubscript{1}} ( 1 - \exp( - m\textsubscript{1} t)) + \frac{n}{m\textsubscript{2}} ( 1 - \exp( - m
\textsubscript{2} t)) - nt 
\end{equation}

This is a transcendental equation which can be solved numerically: the duration of the implied overshoot occurs for the (non-trivial) root of $ \Delta C_{ex} = 0 $, while the maximum carbon budget overshoot - and, approximately, the maximum temperature overshoot - occurs when emissions are zero. Therefore, in a peak-shaving scenario, maximum SRM deployment is only reached when net zero emissions are also reached. For a numerical and graphical exploration of the double-exponential model, as well as a comparison to the quadratic approximation, see Appendix A. 

\subsection{Path-dependence of the TCRE} 

The preceding analysis has assumed that the TCRE is constant for both negative and positive emissions. However, more detailed modelling of the carbon-climate cycle suggests that the TCRE is path-dependent, owing to the lagged thermal response of the ocean \cite{zickfeld2016proportionality}, which means that negative emissions are less effective at cooling than emissions are at warming, at least in the short term. 

\medskip

This non-linearity is most prominent when the lagged ocean response is large, which occurs when the carbon budget overshoot is large. For example, Zickfeld et al. (2016) modelled an emissions trajectory where CO\textsubscript{2} concentrations were increased by 1\% per annum to 4x preindustrial levels and then decreased in the same way. They found that the TCRE was approximately 1.59K in the ascending branch and 1.48K in the descending branch, i.e. 7\% lower.  In scenarios with smaller total cumulative emissions, the difference between the ascending and descending branch was smaller (or even of the opposite sign, implying negative emissions appeared to be slightly more effective at cooling than positive emissions were at warming.) This implies that the analytical estimates above are likely to be slight underestimates for the real-world commitment to peak-shaving; for more detailed analysis of this, see Appendix B.

\section{Example: The SSP-Overshoot Scenario}

We can validate these approximations by considering concrete peak-shaving scenarios. Tilmes et al. (2016) \cite{tilmes2016climate} used the Community Earth System Model (CESM) to simulate the SSP5-3.4-OS scenario \cite{gmd-2019-222}. In this scenario, emissions follow the RCP8.5 pathway until 2040. Then, emissions follow an idealised pathway, declining to become net-negative by the end of the century. The mitigation rate of -1.2Gt CO\textsubscript{2}/yr/yr, as well as the floor of negative emissions at -18.5GtCO\textsubscript{2}/yr, are both drawn from the Integrated Assessment Modelling literature: specifically, the Shared Socioeconomic Pathway database \cite{Riahi2017}, representing the maximum decarbonisation and negative emissions rates respectively. This pathway therefore represents a peak-shaving scenario at the upper end of what is considered feasible in the current scenario literature: there is a significant exceedance of temperature targets due to following RCP8.5 to 2040, then rapid decarbonisation and large-scale negative emissions. Tilmes et al. simulate SRM by using stratospheric aerosol injections to reduce temperature anomalies to 2K from this pathway.

\medskip

In CESM, which has an equilibrium climate sensitivity of 4K, this pathway results in temperature anomalies which peak at 3K above the preindustrial. The warming rate in CESM approaches 0.05K/year at RCP8.5 2040 when its peak-shaving begins (compare to 0.045K/year in the CMIP5 ensemble for RCP8.5 2040) \cite{risbey2017transient}. The cooling rate during the sustained period of negative emissions is just over 0.01K/year.

\medskip

The analytical approximation above therefore suggests a deployment time for SRM to “peak-shave” from 3K to 2K as around 40 years while temperatures rise and around 100 years as they decline. CESM's explicit simulation of this scenario found that geoengineering is deployed for 130 years to shave the peak below 2C. 

\medskip

For a given emissions trajectory, uncertainties around the climate sensitivity affect the overshoot trajectory. We use the FaIR simple climate model \cite{smith2018FaIR} to sample the IPCC AR5 “likely” range for climate sensitivity \cite{stocker2013climate}. Non-CO\textsubscript{2} emissions are assumed to follow a 1.5C-compatible mitigation pathway from 2020, stabilising at a constant radiative forcing of $ 0.45W/m^{2} $ in the long term. In these versions of the scenario, the negative emissions are switched off when temperatures reach 1.5C. 

\medskip

The results, for the standard overshoot parameters with mitigation beginning in 2020 and 2040, are illustrated in Figure 1. The uncertainty in the emissions and cumulative emissions pathways illustrated corresponds to the 67\% confidence interval for climate sensitivity sampled in FaIR, as the time for which negative emissions are switched off and 1.5C is reached depends on the TCRE. Note that for the low-end climate sensitivity in the AR5 range, no negative emissions are required when rapid mitigation begins in 2020, as the 1.5C threshold is not exceeded.

\begin{figure}[!h]
  \includegraphics[width=\linewidth]{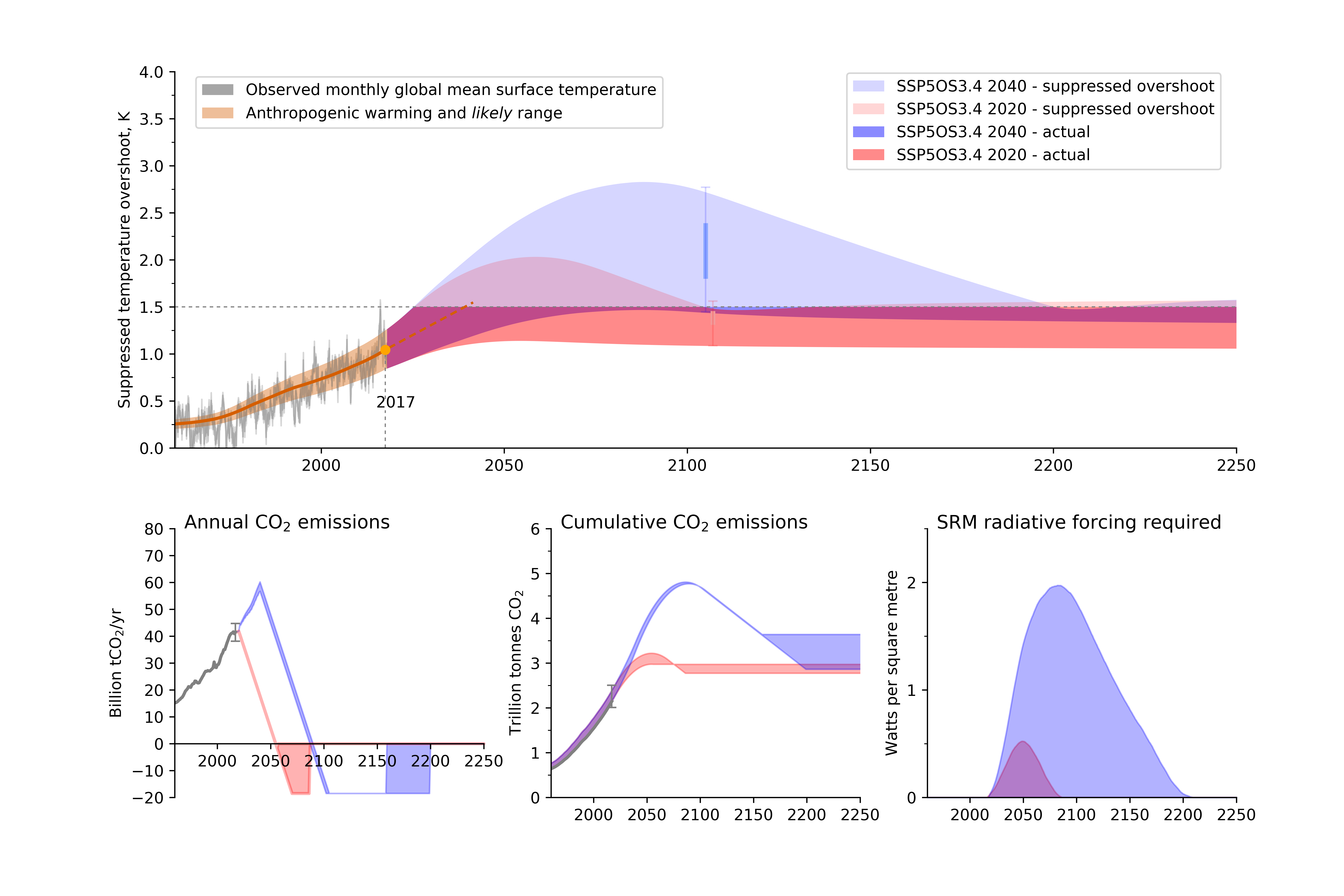}
  \caption{Variants of the SSP5-34-OS scenario, which follows the RCP8.5 scenario with rapid mitigation beginning in 2020 and 2040 respectively. The grey line corresponds to observed temperatures. The suppressed temperature overshoot is shaded with a greater transparency, while the actual pathway temperatures would be expected to follow with SRM is shaded with less transparency. The necessary radiative forcing from SRM to keep temperatures to below 1.5C in these overshoot scenarios is calculated and plotted in the lower right-hand graph. Uncertainty in the temperature response and radiative forcing required, as well as the emissions pathways, corresponds to the 66\% confidence interval for the TCRE from IPCC's AR5 and is plotted with shading. In some cases, this 66\% interval intersects the axis. For example, in the case where mitigation begins in 2020, and the TCRE is at the lower end of estimates, no SRM or negative emissions are required to stay below 1.5C.}
  \label{fig:2020vs2040_SRM.png}
\end{figure}

\newpage

\section{Variants on the overshoot scenario}

To illustrate these points, we consider variants of the idealised SSP5-34-OS case where emissions follow RCP8.5 until a given year (year\_peak), then decline and go negative with an average mitigation rate (mit\_rate), eventually reaching a negative emissions floor (neg\_floor). We randomly sample 5000 sets of year\_peak, mit\_rate, neg\_floor and parameters at from ``feasible" ranges, and use these scenarios to drive FaIR in CO\textsubscript{2}-only mode.  Note that FaIR includes a sufficiently complex carbon cycle that its TCRE is path-dependent, as discussed in Appendix B. 

\medskip

The mitigation rates are sampled from the interval [0.1, 2.3] GtCO\textsubscript{2}/yr/yr. 2.3GtCO\textsubscript{2}/yr/yr is the maximum mitigation rate achieved at any point in the ambitious Low Energy Demand scenario \cite{grubler2018low}. The negative emissions rate is sampled from the range [3.5, 30.0] GtCO\textsubscript{2}/yr. While the lower end of this range assumes that sustained large-scale BECCS and DAC are unavailable, and is determined by the land-use change negative emissions from the Low Energy Demand scenario, the high end of this range is well beyond the maximum considered feasible for BECCS by any 1.5C-compliant scenario, and substantially higher than the largest negative emissions of 18.5GtCO\textsubscript{2}/yr in the scenario database.

\begin{figure}[!h]
  \includegraphics[width=\linewidth]{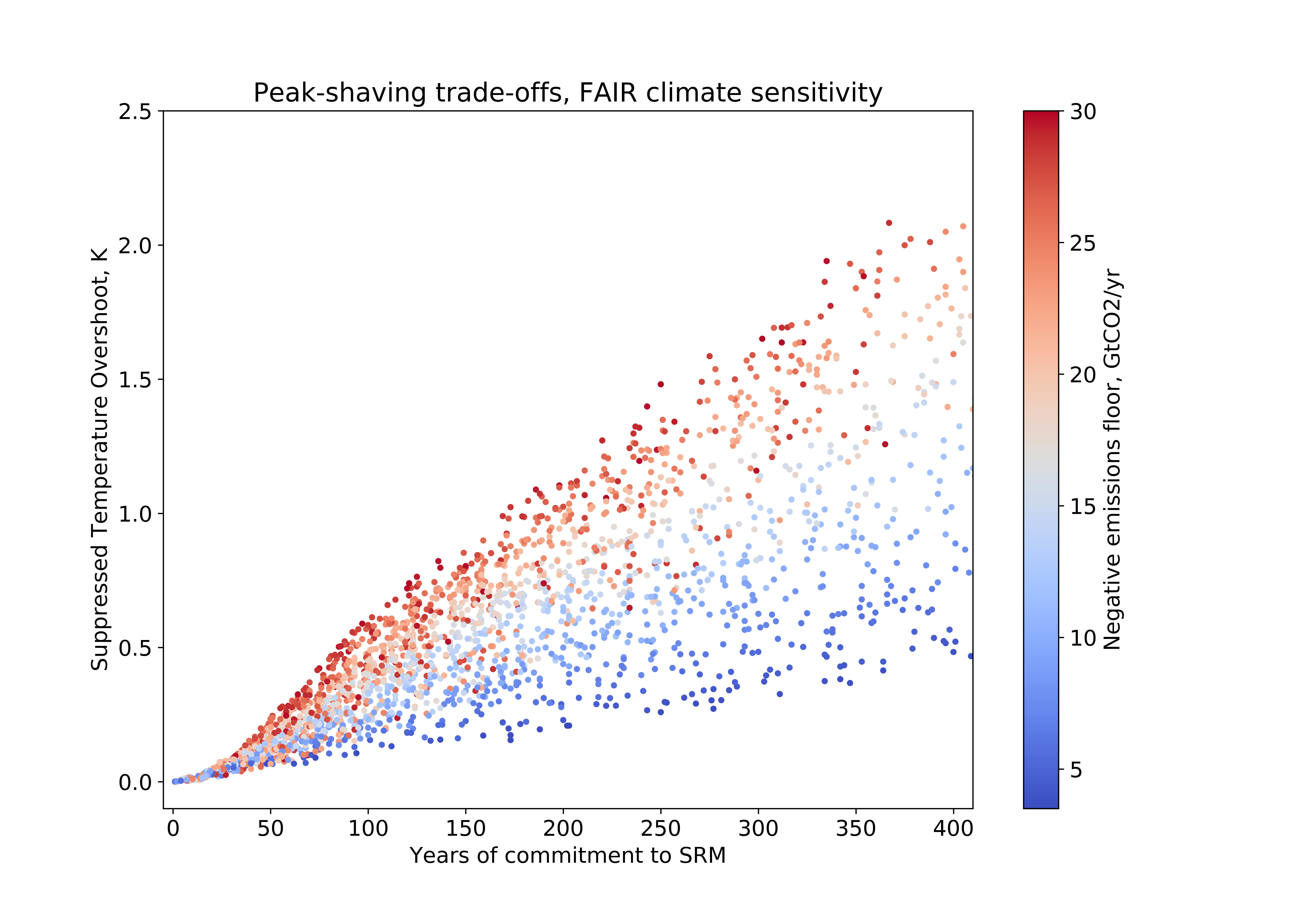}
  \caption{Trade-off between years of committment to solar radiation management and the suppressed temperature overshoot in a range of idealised overshoot and negative emissions scenarios, under the FaIR default ECS and TCR of 2.86K and 1.53K respectively. The idealised scenarios are simulated in FaIR, and are generated by randomly selecting mitigation rate, maximum negative emissions rate, and the year that mitigation begins from ranges of feasible values. There is clearly a strong trade-off between the suppressed temperature overshoot and the years of commitment to SRM in any of the scenarios. Even with a very rapid mitigation rate and a high negative emissions floor, scenarios that suppress more than 0.5K of temperature overshoot generally require around a century of commitment to SRM.}
  \label{fig:tradeoff_fig_1_sto.png}
\end{figure}

\medskip

Realmonte et al. (2019) compared the use of DAC across two integrated assessment models, considering a maximum deployment by 2100 of 30GtCO\textsubscript{2}/yr and a maximum scale-up rate of 1.5 GtCO\textsubscript{2}/yr/yr \cite{realmonte2019inter}. This does not imply that the most extreme scenarios sampled here are feasible - for example, the rapid mitigation under the LED scenario may not be compatible with the large-scale expansion in energy required for DAC on a 30GtCO\textsubscript{2}/yr scale. If large-scale direct air capture is the predominant form of negative emissions, then constraints are likely to be set by societal willingness-to-pay, energy availability, and the speed that the technology can be scaled up and deployed. Realmonte et al. explore some of the implications of deploying DAC at this scale - which would also apply to corresponding peak-shaving scenarios.

\medskip

The year that mitigation starts is sampled between [2020, 2060] – CO\textsubscript{2} emissions follow RCP8.5 prior to this year. Scenarios that do not overshoot 1.5K, or which do not decline below 1.5K during the 750-year integration time of the model, are filtered out. Under these scenarios, the trade-offs between the suppressed temperature overshoot and the years of commitment to SRM are illustrated in Figure 2.  

\medskip

This illustrates the trade-off between the duration of the commitment and the magnitude of the suppressed temperature overshoot under a range of different scenarios. Even if the most rapid mitigation and large-scale negative emissions are assumed, shaving a peak of 2C to 1.5C requires a deployment in excess of 80 years. If negative emissions are not available at sufficient scale ($>2-3$GtCO\textsubscript{2}/yr), shaving a substantial ($>$0.5K peak) becomes impossible on the ~450-year timescale considered here. 

\medskip

\begin{figure}[!h]
  \includegraphics[width=\linewidth]{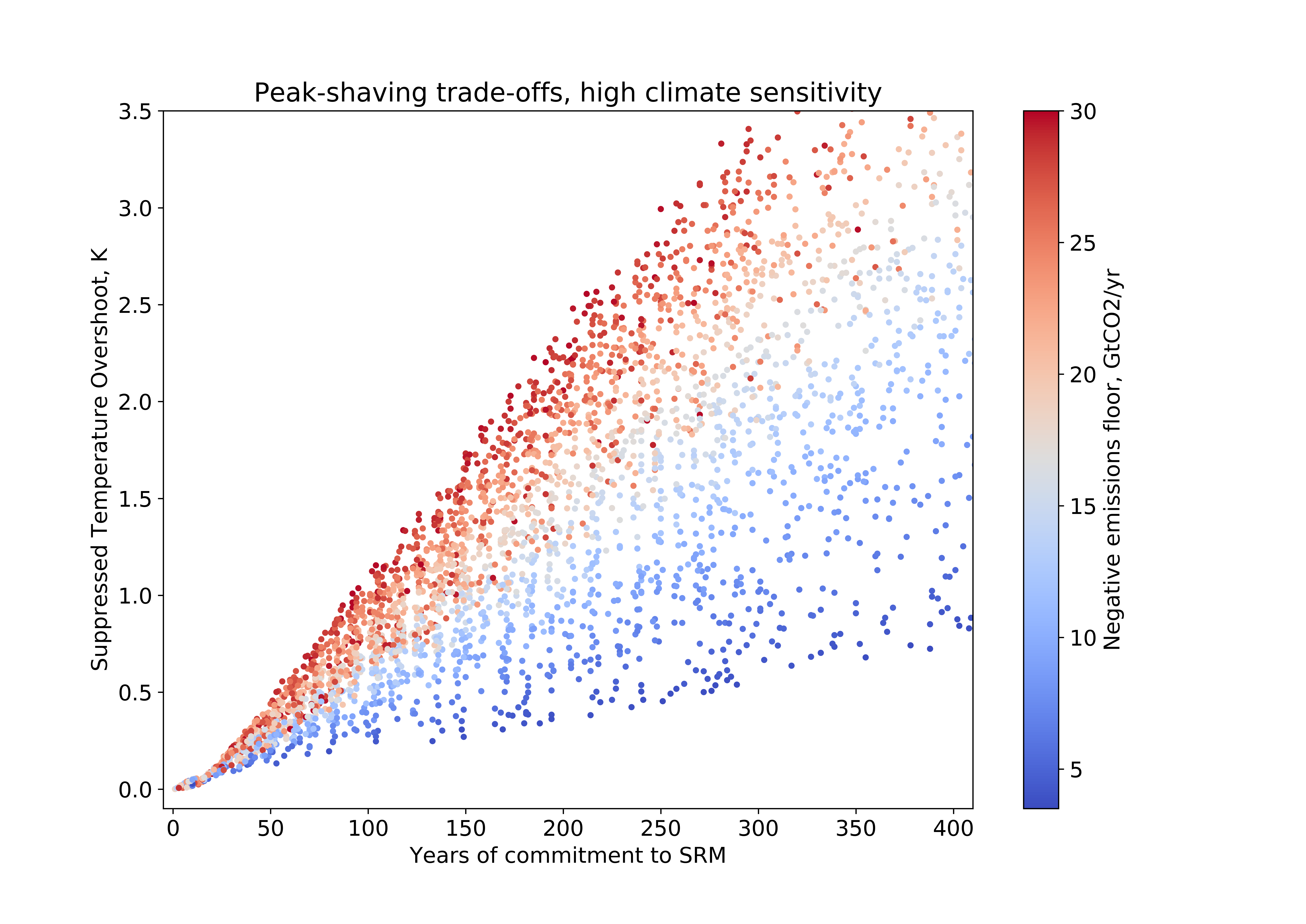}
  \caption{As Figure 2, but with the UKESM1 ECS and TCR of 5.4K and 2.7K respectively, at the higher end of any estimate for climate sensitivity. This illustrates that the trade-off between geoengineering duration and the suppressed temperature overshoot is arguably more favourable when climate sensitivity is high. Because the TCRE is high, small CO\textsubscript{2} excursions lead to significant rises in temperature. This means that the CO\textsubscript{2} corresponding to a temperature rise of 0.5K could be removed in decades, rather than around a century as in the low climate sensitivity case.}
  \label{fig:tradeoff_fig_2.png}
\end{figure}

\medskip

This is dependent on climate sensitivity, as illustrated by Figure 3. With the UKESM climate sensitivity \cite{doi:10.1029/2019MS001739},  temperature overshoots are larger and occur more quickly for the same emissions trajectory. Higher climate sensitivites mean smaller carbon budget overshoots correspond to larger temperature overshoots. Given that the duration of SRM deployment is equivalent to the time spent overshooting the carbon budget, added to the time required to remove the carbon budget overshoot, higher climate sensitivity implies shorter overshoot commitment times to avoid the same level of warming. 

\medskip

In other words, if climate sensitivity is low, then it is difficult to imagine even a contrived scenario where SRM is both useful - suppressing a significant amount of warming, say half a degree - and of short duration, over within a few decades. Deployments of SRM that suppress a significant degree of warming would need to be sustained for centuries, due to the significant additional CO\textsubscript{2} in the atmosphere for that degree of warming, which must be removed. This tradeoff, however, is less dramatic when the climate sensitivity is higher. 

\medskip

The requirement to remove the entire overshoot of the carbon budget before temperatures will decline to below the threshold results in a high sensitivity of the duration of the commitment to any ``floor'' in achievable negative emissions, especially for large overshoots. Table 1 shows how carbon budget overshoots depend on the TCRE and the temperature overshoot. If geoengineering is to be deployed as part of a ``peak-shaving" strategy, then the cost of removing carbon dioxide on this scale must be factored in to the cost of that strategy.

\medskip

\begin{table}[!h]

\def\arraystretch{2}
\centering

\begin{tabular}{|l|l|l|l|l|}
\hline
\textbf{TCRE $\downarrow$ /Overshoot $\rightarrow$} &   0.5K   &   1K &   1.5K & 2K   \\ \hline
0.92K (17\%)            & 543  & 1086 & 1629 & 2172 \\ \hline
1.26K (33\%)            & 396  & 793  & 1189 & 1586 \\ \hline
1.5K (50\%)             & 333  & 666  & 999  & 1332 \\ \hline
1.71K (67\%)            & 292  & 584  & 876  & 1168 \\ \hline
2.05K (83\%)            & 244  & 487  & 731  & 975  \\ \hline
\end{tabular}

\caption{Implied carbon budget overshoots (GtCO\textsubscript{2}) for various levels of temperature overshoot and TCREs from the AR5 WG1 distribution. These correspond to the cumulative net-negative emissions required to return to 1.5K in each overshoot case.} 
\end{table}

\medskip

\section{Conclusion}

This exercise is not intended to span the entire conceivable scenario space of SRM deployment, nor to evaluate the circumstances under which SRM deployment might result in improved climate outcomes, at least for some. Instead, we aim to illustrate tradeoffs inherent in “peak-shaving” scenarios: between the implied commitment time of SRM deployment and the suppressed temperature overshoot, as well as to outline implications for the mitigation rate and scale of negative emissions that are required to ensure the deployment is limited to a particular timescale. 

\medskip

This analysis is arguably conservative, both for the timescales and negative emissions commitment concerned. It assumes that the carbon budget is exceeded rapidly, with high emissions, and that mitigation begins immediately after emissions peak, which both reduce the carbon budget and temperature overshoot compared to scenarios where changes in global emissions occur more slowly.  Here, we consider only the scale of global net-negative emissions. However, additional negative emissions will likely be required even to reach net-zero emissions \cite{Anderson182}, so the actual requirement for negative emissions will be higher. 

\medskip

This analysis also allows for the largest mitigation and negative emissions rates found in the ambitious 1.5C-compliant literature. In a world where SRM is deployed, assuming it is effective in small doses at reducing climate risks \cite{irvine2019halving}, the perceived urgency of reducing emissions and deploying negative emissions at scale may be reduced, a form of ``mitigation deterrence".

\medskip 

Finally, as mentioned, the analytic section assumes that the TCRE is the same for both positive and negative emissions, while more detailed carbon-cycle and climate modelling suggests that the TCRE is path-dependent and that negative emissions are less effective at reducing temperature than positive emissions are at increasing temperature, especially when the cumulative emissions are large. Where this is important, it is likely to increase the timescales considered, unless the realised warming fraction is high. 

\medskip

Unless climate sensitivity is at the high end of estimates, deploying SRM to shave a substantial peak over 1.5C – say, stabilising temperatures at 1.5C when they might have peaked at 2C – is likely to require a commitment to SRM on the order of a century, and to large-scale negative emissions, cumulatively at least 350GtCO\textsubscript{2}. We find that, under median estimates of climate sensitivity, deployments of SRM that are complete within a few decades are unlikely to shift global mean temperatures a great deal beyond multidecadal variability. Cost-benefit analyses that assume SRM deployment is not indefinite must consider the cost and feasibility of cumulative negative emissions equivalent to the exceedance of the carbon budget, as well as the timescales required to draw these emissions down.

\medskip

During the COVID-19 pandemic, to avoid loss of life, many of the world's governments have been implemented severe social distancing measures in order to ``flatten the curve" of infection. We see an analogy between these two situations. The R-rate during the uncontrolled initial phase of the epidemic is analagous to our emissions before net-zero is reached, and the rate of negative emissions we can attain is analagous to the R-rate in lockdown. The curves of new case numbers in countries that implemented lockdowns resembled the asymmetric parabolas of section 2.2, as in general the R during the uncontrolled epidemic phase was larger than the reciprocal of the R during lockdown. Therefore, each additional week of delay prior to lockdown required several weeks of commitment to lockdown to return infection numbers to their previous levels, as well as entailing a greater overall burden in terms of death and economic damages. Countries that were able to implement early lockdowns, prior to substantial spread of the disease, were generally able to adopt less stringent, less expensive, and shorter-duration social distancing measures, while experiencing a lower rate of mortality. Similarly, the more rapidly mitigation takes place, the shorter any long-term committment to costly geoengineering measures would be.

\medskip

In both cases, the inescapable conclusion is that the optimal scenario involves early, swift and decisive action to tackle the root of the problem, which obviates the need for a long-term commitment to a costly suppression policy. When this becomes impossible, the trade-offs illustrated in this paper must be confronted.

\newpage

\appendix

\section{Double-exponential model}

Considering the double-exponential idealised model for emissions, with negative emissions that increase towards a floor and emissions that decay simply with mitigation, emissions after mitigation begins can be expressed as:

\begin{equation}
E(t) =  C_{max} \exp(-  m\textsubscript{1} t) - n ( 1 - \exp( - m\textsubscript{2} t)) 
\end{equation}

where

\begin{equation}
m\textsubscript{1,2} = {\ln} (2)/t\textsubscript{1,2} 
\end{equation}

and t\textsubscript{1,2} are the timescales for halving (gross) emissions and scaling up the deployment of negative emissions to half of its maximum value, respectively. 

\begin{figure}[!h]
  \includegraphics[width=\linewidth]{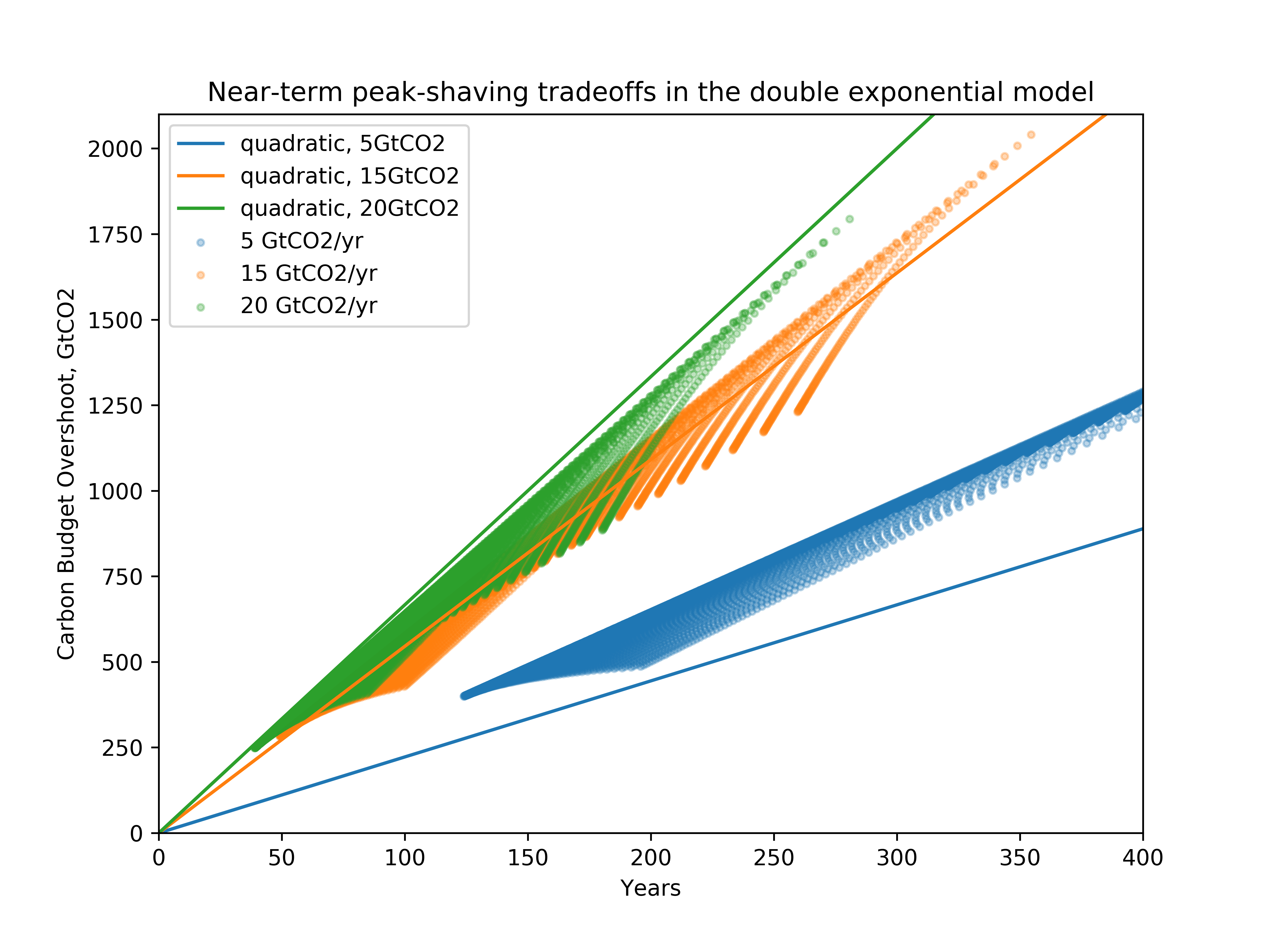}
  \caption{Illustration of the tradeoff between duration and height of overshoot in the double-exponential emissions model, taking TCRE to be 1.4K/GtC as in the FaIR model. Mitigation and negative emissions deployment rates are both chosen in the interval [1, 7]\% and three negative emissions floors are sampled. The quadratic model is approximated as in Equation 6, assuming that the emissions when the threshold is crossed are the peak of emissions on the way up, and the negative emissions floor on the way down.}
  \label{fig:2020vs2040_SRM.png}
\end{figure}

\medskip

As an example, one could consider emissions peaking close to their present value of ~40GtCO\textsubscript{2}/year, then declining (through conventional mitigation) at a given annual average percentage while negative emissions are scaled up to their maximum value. Different examples illustrate the trade-off between the implied duration of and height of the peak. For this example, we choose mitigation rates between a maximum of 7\% per annum, which corresponds to halving emissions in just under a decade, and 0.5\%, which corresponds to halving emissions in around 140 years. This is illustrated in Figure 4. Note the tradeoff between the duration of geoengineering and the height of the peak shaved in all scenarios. The quadratic approximation, where mitigation occurs at a constant rate in terms of gigatonnes of carbon, is also illustrated and its gradient depends entirely on the rate of negative emissions, rather than the mitigation and deployment rates. 

\medskip 

It might be argued that the best case for peak-shaving is a scenario where conventional mitigation is slow or expensive, but negative emissions can eventually be deployed more rapidly - for example, if very cheap energy becomes available at scale later in this century. These scenarios are sampled here, but even in this extreme case shaving a peak of 0.5K involves 70+ years of geoengineering.

\newpage

\section{Path-dependence of the TCRE}

Depending on the emissions scenario and the model used, the approximation that global mean temperature is proportional to cumulative emissions - and hence that TCRE is constant and well-defined - can be less accurate. Physically, the concept of a constant TCRE depends on the approximate cancellation of two factors: residual surface warming as the ocean comes into equilibrium with the high-CO\textsubscript{2} atmosphere, and uptake of carbon dioxide by the ocean sink and biosphere \cite{allen2009warming}. Consequently, factors that vary between models - such as the rate of uptake of CO\textsubscript{2} by the oceans and the biosphere, as well as the timescale and magnitude of long-term warming as the deep ocean comes into equilibrium with the atmosphere - can introduce a hysteresis effect in temperature as cumulative emissions are reduced, and hence an apparent time-dependence of the TCRE. 

\begin{figure}[!h]
  \includegraphics[width=\linewidth]{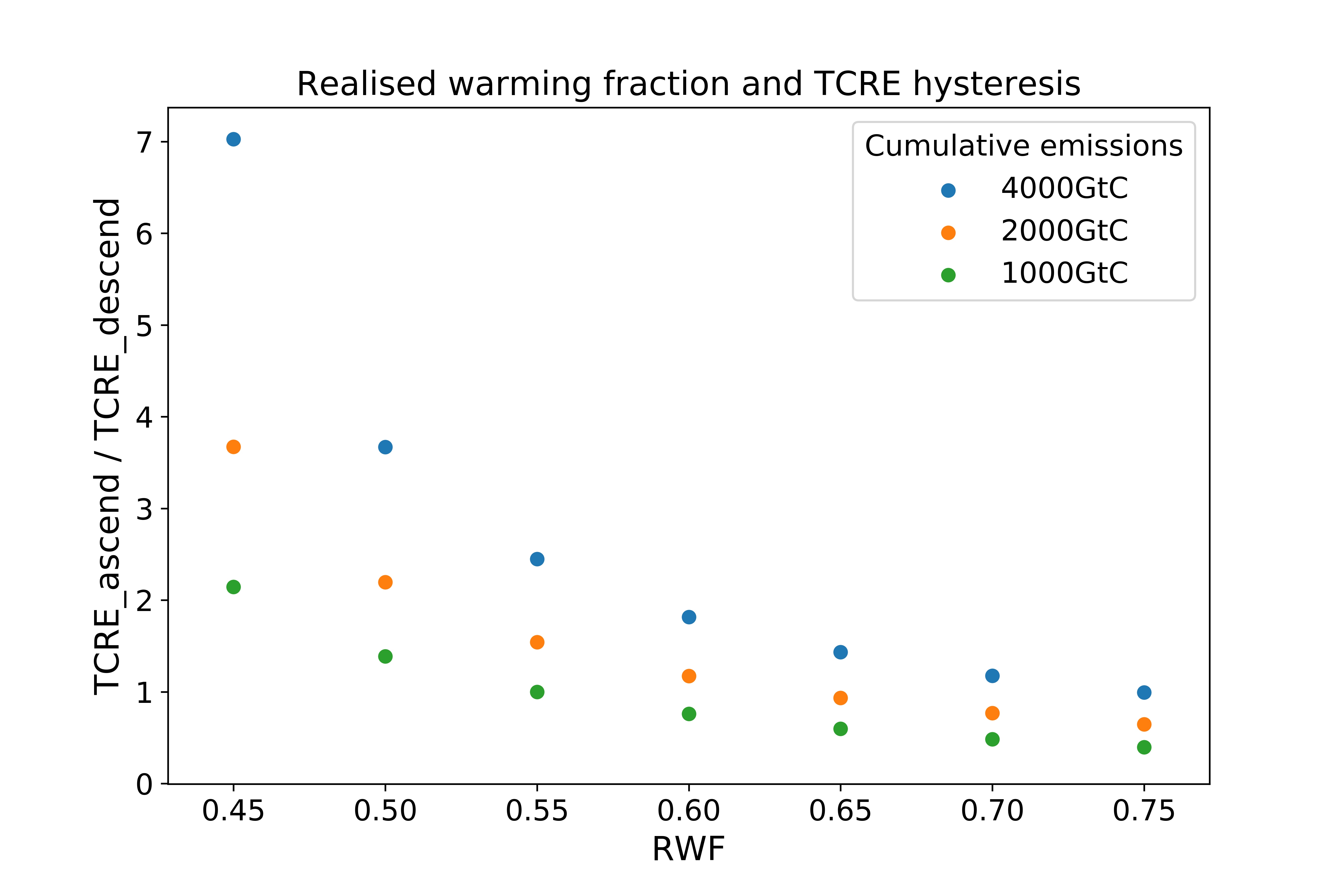}
  \caption{Ratio of TCRE during positive and negative emissions periods for various Realised Warming Fractions (TCR/ECS ratios). This was calculated by simulating simple emissions trajectory where emissions linearly decrease to zero and then become negative at the same rate, until cumulative emissions are zero. Differences both in the total anthropogenic CO\textsubscript{2} excursion and the RWF give rise to different levels of hysteresis. The approximation that the TCRE is constant for removals and additions of CO\textsubscript{2} - which is relied on for the simple algebraic models, but not the FaIR analysis - breaks down for RWF far from the FaIR default of 0.53, or for large anthropogenic CO\textsubscript{2} excursions. Lower RWF or greater CO\textsubscript{2} excursions would result in a greater implied duration of geoengineering to maintain temperatures below a given threshold.}
  \label{fig:TCRE_RWF_hysteresis}
\end{figure}

This behaviour can be modelled with FaIR by using idealised emissions trajectories and varying the TCR/ECS ratio, known as the "Realised Warming Fraction" (RWF). RWFs for the CMIP5 models lie in the range 0.45-0.7, with an average of 0.55: observationally constrained estimates tend to fall in the upper half of this range. \cite{millar2017modified} In these scenarios, emissions linearly decrease from an initial maximum value E to -E over a period of 200 years, resulting in net cumulative emissions of zero. The TCRE is evaluated for the ascending and descending branches by calculating the gradient of the temperature-cumulative emissions curve, then averaging across the periods of positive and negative emissions respectively. We plot cases for three peak cumulative emissions trajectories. The ratio of the TCRE in the ascending and descending branches is plotted against the RWF for several emissions scenarios in Figure 5.

If the RWF is low, or the total carbon perturbation is high, the assumption of linearity is less accurate. In these cases, global mean temperature can continue to rise even as emissions go negative, as the large ``unrealised" warming dominates the effect of declining CO\textsubscript{2} concentrations, reducing the average TCRE across this descending branch. This will result in the analytical formulae of Section 2 underestimating the commitment time to peak-shaving geoengineering. By contrast, if the RWF is high, and the carbon perturbation is small enough, the TCRE ratio can be less than 1: in other words, natural removal of CO\textsubscript{2} by sinks exceeds the effect of long-term warming, which results in faster cooling during the negative emissions branch, as concentrations decline more quickly. 

The non-constant nature of the TCRE is an important caveat for the early, analytical calculations of the trade-off between duration and the height of the peak shaved - especially if the RWF is at the extreme ends of current estimates, or the carbon perturbation is high. However, this effect is small for RWF of around 0.5-0.6, similar to the multi-model CMIP5 average of 0.55 and the FaIR default of 0.53.

\newpage

\acknowledgments

T.H. was funded for his DPhil by the Natural Environment Research Council's Doctoral Training Partnership for Environmental Research. 


%
%

\bibliography{mybib}

\begin{thebibliography}{}

\bibitem [\protect \citeauthoryear {%
Allen%
\ \protect \BOthers {.}}{%
Allen%
\ \protect \BOthers {.}}{%
{\protect \APACyear {2009}}%
}]{%
allen2009warming}
\APACinsertmetastar {%
allen2009warming}%
\begin{APACrefauthors}%
Allen, M\BPBI R.%
, Frame, D\BPBI J.%
, Huntingford, C.%
, Jones, C\BPBI D.%
, Lowe, J\BPBI A.%
, Meinshausen, M.%
\BCBL {}\ \BBA {} Meinshausen, N.%
\end{APACrefauthors}%
\unskip\
\newblock
\APACrefYearMonthDay{2009}{}{}.
\newblock
{\BBOQ}\APACrefatitle {Warming caused by cumulative carbon emissions towards
  the trillionth tonne} {Warming caused by cumulative carbon emissions towards
  the trillionth tonne}.{\BBCQ}
\newblock
\APACjournalVolNumPages{Nature}{458}{7242}{1163}.
\PrintBackRefs{\CurrentBib}

\bibitem [\protect \citeauthoryear {%
Anderson%
\ \BBA {} Peters%
}{%
Anderson%
\ \BBA {} Peters%
}{%
{\protect \APACyear {2016}}%
}]{%
Anderson182}
\APACinsertmetastar {%
Anderson182}%
\begin{APACrefauthors}%
Anderson, K.%
\BCBT {}\ \BBA {} Peters, G.%
\end{APACrefauthors}%
\unskip\
\newblock
\APACrefYearMonthDay{2016}{}{}.
\newblock
{\BBOQ}\APACrefatitle {The trouble with negative emissions} {The trouble with
  negative emissions}.{\BBCQ}
\newblock
\APACjournalVolNumPages{Science}{354}{6309}{182--183}.
\newblock
\begin{APACrefURL} \url{https://science.sciencemag.org/content/354/6309/182}
  \end{APACrefURL}
\newblock
\begin{APACrefDOI} \doi{10.1126/science.aah4567} \end{APACrefDOI}
\PrintBackRefs{\CurrentBib}

\bibitem [\protect \citeauthoryear {%
de Coninck%
\ \protect \BOthers {.}}{%
de Coninck%
\ \protect \BOthers {.}}{%
{\protect \APACyear {2018}}%
}]{%
de2018strengthening}
\APACinsertmetastar {%
de2018strengthening}%
\begin{APACrefauthors}%
de Coninck, H.%
, Revi, A.%
, Babiker, M.%
, Bertoldi, P.%
, Buckeridge, M.%
, Cartwright, A.%
\BDBL {}others%
\end{APACrefauthors}%
\unskip\
\newblock
\APACrefYearMonthDay{2018}{}{}.
\newblock
{\BBOQ}\APACrefatitle {Strengthening and implementing the global response}
  {Strengthening and implementing the global response}.{\BBCQ}
\newblock
\APACjournalVolNumPages{IPCC}{}{}{}.
\PrintBackRefs{\CurrentBib}

\bibitem [\protect \citeauthoryear {%
Fasullo%
\ \protect \BOthers {.}}{%
Fasullo%
\ \protect \BOthers {.}}{%
{\protect \APACyear {2018}}%
}]{%
fasullo2018persistent}
\APACinsertmetastar {%
fasullo2018persistent}%
\begin{APACrefauthors}%
Fasullo, J\BPBI T.%
, Tilmes, S.%
, Richter, J\BPBI H.%
, Kravitz, B.%
, MacMartin, D\BPBI G.%
, Mills, M\BPBI J.%
\BCBL {}\ \BBA {} Simpson, I\BPBI R.%
\end{APACrefauthors}%
\unskip\
\newblock
\APACrefYearMonthDay{2018}{}{}.
\newblock
{\BBOQ}\APACrefatitle {Persistent polar ocean warming in a strategically
  geoengineered climate} {Persistent polar ocean warming in a strategically
  geoengineered climate}.{\BBCQ}
\newblock
\APACjournalVolNumPages{Nature Geoscience}{11}{12}{910}.
\PrintBackRefs{\CurrentBib}

\bibitem [\protect \citeauthoryear {%
Goes%
, Tuana%
\BCBL {}\ \BBA {} Keller%
}{%
Goes%
\ \protect \BOthers {.}}{%
{\protect \APACyear {2011}}%
}]{%
goes2011economics}
\APACinsertmetastar {%
goes2011economics}%
\begin{APACrefauthors}%
Goes, M.%
, Tuana, N.%
\BCBL {}\ \BBA {} Keller, K.%
\end{APACrefauthors}%
\unskip\
\newblock
\APACrefYearMonthDay{2011}{}{}.
\newblock
{\BBOQ}\APACrefatitle {The economics (or lack thereof) of aerosol
  geoengineering} {The economics (or lack thereof) of aerosol
  geoengineering}.{\BBCQ}
\newblock
\APACjournalVolNumPages{Climatic change}{109}{3-4}{719--744}.
\PrintBackRefs{\CurrentBib}

\bibitem [\protect \citeauthoryear {%
Grubler%
\ \protect \BOthers {.}}{%
Grubler%
\ \protect \BOthers {.}}{%
{\protect \APACyear {2018}}%
}]{%
grubler2018low}
\APACinsertmetastar {%
grubler2018low}%
\begin{APACrefauthors}%
Grubler, A.%
, Wilson, C.%
, Bento, N.%
, Boza-Kiss, B.%
, Krey, V.%
, McCollum, D\BPBI L.%
\BDBL {}others%
\end{APACrefauthors}%
\unskip\
\newblock
\APACrefYearMonthDay{2018}{}{}.
\newblock
{\BBOQ}\APACrefatitle {A low energy demand scenario for meeting the 1.5 C
  target and sustainable development goals without negative emission
  technologies} {A low energy demand scenario for meeting the 1.5 c target and
  sustainable development goals without negative emission technologies}.{\BBCQ}
\newblock
\APACjournalVolNumPages{Nature Energy}{3}{6}{515}.
\PrintBackRefs{\CurrentBib}

\bibitem [\protect \citeauthoryear {%
Irvine%
\ \protect \BOthers {.}}{%
Irvine%
\ \protect \BOthers {.}}{%
{\protect \APACyear {2019}}%
}]{%
irvine2019halving}
\APACinsertmetastar {%
irvine2019halving}%
\begin{APACrefauthors}%
Irvine, P.%
, Emanuel, K.%
, He, J.%
, Horowitz, L\BPBI W.%
, Vecchi, G.%
\BCBL {}\ \BBA {} Keith, D.%
\end{APACrefauthors}%
\unskip\
\newblock
\APACrefYearMonthDay{2019}{}{}.
\newblock
{\BBOQ}\APACrefatitle {Halving warming with idealized solar geoengineering
  moderates key climate hazards} {Halving warming with idealized solar
  geoengineering moderates key climate hazards}.{\BBCQ}
\newblock
\APACjournalVolNumPages{Nature Climate Change}{9}{4}{295}.
\PrintBackRefs{\CurrentBib}

\bibitem [\protect \citeauthoryear {%
Jones%
\ \protect \BOthers {.}}{%
Jones%
\ \protect \BOthers {.}}{%
{\protect \APACyear {2013}}%
}]{%
jones2013impact}
\APACinsertmetastar {%
jones2013impact}%
\begin{APACrefauthors}%
Jones, A.%
, Haywood, J\BPBI M.%
, Alterskj{\ae}r, K.%
, Boucher, O.%
, Cole, J\BPBI N.%
, Curry, C\BPBI L.%
\BDBL {}others%
\end{APACrefauthors}%
\unskip\
\newblock
\APACrefYearMonthDay{2013}{}{}.
\newblock
{\BBOQ}\APACrefatitle {The impact of abrupt suspension of solar radiation
  management (termination effect) in experiment G2 of the Geoengineering Model
  Intercomparison Project (GeoMIP)} {The impact of abrupt suspension of solar
  radiation management (termination effect) in experiment g2 of the
  geoengineering model intercomparison project (geomip)}.{\BBCQ}
\newblock
\APACjournalVolNumPages{Journal of Geophysical Research:
  Atmospheres}{118}{17}{9743--9752}.
\PrintBackRefs{\CurrentBib}

\bibitem [\protect \citeauthoryear {%
Keith%
\ \BBA {} Irvine%
}{%
Keith%
\ \BBA {} Irvine%
}{%
{\protect \APACyear {2016}}%
}]{%
keith2016solar}
\APACinsertmetastar {%
keith2016solar}%
\begin{APACrefauthors}%
Keith, D\BPBI W.%
\BCBT {}\ \BBA {} Irvine, P\BPBI J.%
\end{APACrefauthors}%
\unskip\
\newblock
\APACrefYearMonthDay{2016}{}{}.
\newblock
{\BBOQ}\APACrefatitle {Solar geoengineering could substantially reduce climate
  risks—A research hypothesis for the next decade} {Solar geoengineering
  could substantially reduce climate risks—a research hypothesis for the next
  decade}.{\BBCQ}
\newblock
\APACjournalVolNumPages{Earth's Future}{4}{11}{549--559}.
\PrintBackRefs{\CurrentBib}

\bibitem [\protect \citeauthoryear {%
Kleinschmitt%
, Boucher%
\BCBL {}\ \BBA {} Platt%
}{%
Kleinschmitt%
\ \protect \BOthers {.}}{%
{\protect \APACyear {2018}}%
}]{%
kleinschmitt2018sensitivity}
\APACinsertmetastar {%
kleinschmitt2018sensitivity}%
\begin{APACrefauthors}%
Kleinschmitt, C.%
, Boucher, O.%
\BCBL {}\ \BBA {} Platt, U.%
\end{APACrefauthors}%
\unskip\
\newblock
\APACrefYearMonthDay{2018}{}{}.
\newblock
{\BBOQ}\APACrefatitle {Sensitivity of the radiative forcing by stratospheric
  sulfur geoengineering to the amount and strategy of the SO 2 injection
  studied with the LMDZ-S3A model} {Sensitivity of the radiative forcing by
  stratospheric sulfur geoengineering to the amount and strategy of the so 2
  injection studied with the lmdz-s3a model}.{\BBCQ}
\newblock
\APACjournalVolNumPages{Atmospheric Chemistry and Physics}{18}{4}{2769--2786}.
\PrintBackRefs{\CurrentBib}

\bibitem [\protect \citeauthoryear {%
MacMartin%
, Ricke%
\BCBL {}\ \BBA {} Keith%
}{%
MacMartin%
\ \protect \BOthers {.}}{%
{\protect \APACyear {2018}}%
}]{%
macmartin2018solar}
\APACinsertmetastar {%
macmartin2018solar}%
\begin{APACrefauthors}%
MacMartin, D\BPBI G.%
, Ricke, K\BPBI L.%
\BCBL {}\ \BBA {} Keith, D\BPBI W.%
\end{APACrefauthors}%
\unskip\
\newblock
\APACrefYearMonthDay{2018}{}{}.
\newblock
{\BBOQ}\APACrefatitle {Solar geoengineering as part of an overall strategy for
  meeting the 1.5 C Paris target} {Solar geoengineering as part of an overall
  strategy for meeting the 1.5 c paris target}.{\BBCQ}
\newblock
\APACjournalVolNumPages{Philosophical Transactions of the Royal Society A:
  Mathematical, Physical and Engineering Sciences}{376}{2119}{20160454}.
\PrintBackRefs{\CurrentBib}

\bibitem [\protect \citeauthoryear {%
Meinshausen%
\ \protect \BOthers {.}}{%
Meinshausen%
\ \protect \BOthers {.}}{%
{\protect \APACyear {2019}}%
}]{%
gmd-2019-222}
\APACinsertmetastar {%
gmd-2019-222}%
\begin{APACrefauthors}%
Meinshausen, M.%
, Nicholls, Z.%
, Lewis, J.%
, Gidden, M\BPBI J.%
, Vogel, E.%
, Freund, M.%
\BDBL {}Wang, H\BPBI J.%
\end{APACrefauthors}%
\unskip\
\newblock
\APACrefYearMonthDay{2019}{}{}.
\newblock
{\BBOQ}\APACrefatitle {The SSP greenhouse gas concentrations and their
  extensions to 2500} {The ssp greenhouse gas concentrations and their
  extensions to 2500}.{\BBCQ}
\newblock
\APACjournalVolNumPages{Geoscientific Model Development
  Discussions}{2019}{}{1--77}.
\newblock
\begin{APACrefURL} \url{https://www.geosci-model-dev-discuss.net/gmd-2019-222/}
  \end{APACrefURL}
\newblock
\begin{APACrefDOI} \doi{10.5194/gmd-2019-222} \end{APACrefDOI}
\PrintBackRefs{\CurrentBib}

\bibitem [\protect \citeauthoryear {%
Millar%
, Nicholls%
, Friedlingstein%
\BCBL {}\ \BBA {} Allen%
}{%
Millar%
\ \protect \BOthers {.}}{%
{\protect \APACyear {2017}}%
}]{%
millar2017modified}
\APACinsertmetastar {%
millar2017modified}%
\begin{APACrefauthors}%
Millar, R\BPBI J.%
, Nicholls, Z\BPBI R.%
, Friedlingstein, P.%
\BCBL {}\ \BBA {} Allen, M\BPBI R.%
\end{APACrefauthors}%
\unskip\
\newblock
\APACrefYearMonthDay{2017}{}{}.
\newblock
{\BBOQ}\APACrefatitle {A modified impulse-response representation of the global
  near-surface air temperature and atmospheric concentration response to carbon
  dioxide emissions} {A modified impulse-response representation of the global
  near-surface air temperature and atmospheric concentration response to carbon
  dioxide emissions}.{\BBCQ}
\newblock
\APACjournalVolNumPages{Atmospheric Chemistry and Physics}{17}{11}{7213--7228}.
\PrintBackRefs{\CurrentBib}

\bibitem [\protect \citeauthoryear {%
Moreno-Cruz%
, Ricke%
\BCBL {}\ \BBA {} Keith%
}{%
Moreno-Cruz%
\ \protect \BOthers {.}}{%
{\protect \APACyear {2012}}%
}]{%
moreno2012simple}
\APACinsertmetastar {%
moreno2012simple}%
\begin{APACrefauthors}%
Moreno-Cruz, J\BPBI B.%
, Ricke, K\BPBI L.%
\BCBL {}\ \BBA {} Keith, D\BPBI W.%
\end{APACrefauthors}%
\unskip\
\newblock
\APACrefYearMonthDay{2012}{}{}.
\newblock
{\BBOQ}\APACrefatitle {A simple model to account for regional inequalities in
  the effectiveness of solar radiation management} {A simple model to account
  for regional inequalities in the effectiveness of solar radiation
  management}.{\BBCQ}
\newblock
\APACjournalVolNumPages{Climatic change}{110}{3-4}{649--668}.
\PrintBackRefs{\CurrentBib}

\bibitem [\protect \citeauthoryear {%
Niemeier%
\ \BBA {} Timmreck%
}{%
Niemeier%
\ \BBA {} Timmreck%
}{%
{\protect \APACyear {2015}}%
}]{%
niemeier2015limit}
\APACinsertmetastar {%
niemeier2015limit}%
\begin{APACrefauthors}%
Niemeier, U.%
\BCBT {}\ \BBA {} Timmreck, C.%
\end{APACrefauthors}%
\unskip\
\newblock
\APACrefYearMonthDay{2015}{}{}.
\newblock
{\BBOQ}\APACrefatitle {What is the limit of climate engineering by
  stratospheric injection of SO 2?} {What is the limit of climate engineering
  by stratospheric injection of so 2?}{\BBCQ}
\newblock
\APACjournalVolNumPages{Atmospheric Chemistry and Physics}{15}{16}{9129--9141}.
\PrintBackRefs{\CurrentBib}

\bibitem [\protect \citeauthoryear {%
Realmonte%
\ \protect \BOthers {.}}{%
Realmonte%
\ \protect \BOthers {.}}{%
{\protect \APACyear {2019}}%
}]{%
realmonte2019inter}
\APACinsertmetastar {%
realmonte2019inter}%
\begin{APACrefauthors}%
Realmonte, G.%
, Drouet, L.%
, Gambhir, A.%
, Glynn, J.%
, Hawkes, A.%
, K{\"o}berle, A\BPBI C.%
\BCBL {}\ \BBA {} Tavoni, M.%
\end{APACrefauthors}%
\unskip\
\newblock
\APACrefYearMonthDay{2019}{}{}.
\newblock
{\BBOQ}\APACrefatitle {An inter-model assessment of the role of direct air
  capture in deep mitigation pathways} {An inter-model assessment of the role
  of direct air capture in deep mitigation pathways}.{\BBCQ}
\newblock
\APACjournalVolNumPages{Nature communications}{10}{1}{3277}.
\PrintBackRefs{\CurrentBib}

\bibitem [\protect \citeauthoryear {%
Riahi%
\ \protect \BOthers {.}}{%
Riahi%
\ \protect \BOthers {.}}{%
{\protect \APACyear {2017}}%
}]{%
Riahi2017}
\APACinsertmetastar {%
Riahi2017}%
\begin{APACrefauthors}%
Riahi, K.%
, van Vuuren, D\BPBI P.%
, Kriegler, E.%
, Edmonds, J.%
, O'Neill, B\BPBI C.%
, Fujimori, S.%
\BDBL {}Tavoni, M.%
\end{APACrefauthors}%
\unskip\
\newblock
\APACrefYearMonthDay{2017}{jan}{}.
\newblock
{\BBOQ}\APACrefatitle {The Shared Socioeconomic Pathways and their energy, land
  use, and greenhouse gas emissions implications: An overview} {The shared
  socioeconomic pathways and their energy, land use, and greenhouse gas
  emissions implications: An overview}.{\BBCQ}
\newblock
\APACjournalVolNumPages{Global Environmental Change}{42}{}{153--168}.
\newblock
\begin{APACrefDOI} \doi{10.1016/j.gloenvcha.2016.05.009} \end{APACrefDOI}
\PrintBackRefs{\CurrentBib}

\bibitem [\protect \citeauthoryear {%
Risbey%
, Grose%
, Monselesan%
, O'Kane%
\BCBL {}\ \BBA {} Lewandowsky%
}{%
Risbey%
\ \protect \BOthers {.}}{%
{\protect \APACyear {2017}}%
}]{%
risbey2017transient}
\APACinsertmetastar {%
risbey2017transient}%
\begin{APACrefauthors}%
Risbey, J\BPBI S.%
, Grose, M\BPBI R.%
, Monselesan, D\BPBI P.%
, O'Kane, T\BPBI J.%
\BCBL {}\ \BBA {} Lewandowsky, S.%
\end{APACrefauthors}%
\unskip\
\newblock
\APACrefYearMonthDay{2017}{}{}.
\newblock
{\BBOQ}\APACrefatitle {Transient response of the global mean warming rate and
  its spatial variation} {Transient response of the global mean warming rate
  and its spatial variation}.{\BBCQ}
\newblock
\APACjournalVolNumPages{Weather and climate extremes}{18}{}{55--64}.
\PrintBackRefs{\CurrentBib}

\bibitem [\protect \citeauthoryear {%
Rockstr{\"o}m%
\ \protect \BOthers {.}}{%
Rockstr{\"o}m%
\ \protect \BOthers {.}}{%
{\protect \APACyear {2017}}%
}]{%
Rockstrom1269}
\APACinsertmetastar {%
Rockstrom1269}%
\begin{APACrefauthors}%
Rockstr{\"o}m, J.%
, Gaffney, O.%
, Rogelj, J.%
, Meinshausen, M.%
, Nakicenovic, N.%
\BCBL {}\ \BBA {} Schellnhuber, H\BPBI J.%
\end{APACrefauthors}%
\unskip\
\newblock
\APACrefYearMonthDay{2017}{}{}.
\newblock
{\BBOQ}\APACrefatitle {A roadmap for rapid decarbonization} {A roadmap for
  rapid decarbonization}.{\BBCQ}
\newblock
\APACjournalVolNumPages{Science}{355}{6331}{1269--1271}.
\newblock
\begin{APACrefURL} \url{https://science.sciencemag.org/content/355/6331/1269}
  \end{APACrefURL}
\newblock
\begin{APACrefDOI} \doi{10.1126/science.aah3443} \end{APACrefDOI}
\PrintBackRefs{\CurrentBib}

\bibitem [\protect \citeauthoryear {%
Sellar%
\ \protect \BOthers {.}}{%
Sellar%
\ \protect \BOthers {.}}{%
{\protect \APACyear {2019}}%
}]{%
doi:10.1029/2019MS001739}
\APACinsertmetastar {%
doi:10.1029/2019MS001739}%
\begin{APACrefauthors}%
Sellar, A\BPBI A.%
, Jones, C\BPBI G.%
, Mulcahy, J.%
, Tang, Y.%
, Yool, A.%
, Wiltshire, A.%
\BDBL {}Zerroukat, M.%
\end{APACrefauthors}%
\unskip\
\newblock
\APACrefYearMonthDay{2019}{}{}.
\newblock
{\BBOQ}\APACrefatitle {UKESM1: Description and evaluation of the UK Earth
  System Model} {Ukesm1: Description and evaluation of the uk earth system
  model}.{\BBCQ}
\newblock
\APACjournalVolNumPages{Journal of Advances in Modeling Earth
  Systems}{0}{ja}{}.
\newblock
\begin{APACrefURL}
  \url{https://agupubs.onlinelibrary.wiley.com/doi/abs/10.1029/2019MS001739}
  \end{APACrefURL}
\newblock
\begin{APACrefDOI} \doi{10.1029/2019MS001739} \end{APACrefDOI}
\PrintBackRefs{\CurrentBib}

\bibitem [\protect \citeauthoryear {%
C\BPBI J.~Smith%
\ \protect \BOthers {.}}{%
C\BPBI J.~Smith%
\ \protect \BOthers {.}}{%
{\protect \APACyear {2018}}%
}]{%
smith2018FaIR}
\APACinsertmetastar {%
smith2018FaIR}%
\begin{APACrefauthors}%
Smith, C\BPBI J.%
, Forster, P\BPBI M.%
, Allen, M.%
, Leach, N.%
, Millar, R\BPBI J.%
, Passerello, G\BPBI A.%
\BCBL {}\ \BBA {} Regayre, L\BPBI A.%
\end{APACrefauthors}%
\unskip\
\newblock
\APACrefYearMonthDay{2018}{}{}.
\newblock
{\BBOQ}\APACrefatitle {FAIR v1. 3: A simple emissions-based impulse response
  and carbon cycle model} {Fair v1. 3: A simple emissions-based impulse
  response and carbon cycle model}.{\BBCQ}
\newblock
\APACjournalVolNumPages{Geoscientific Model Development}{11}{6}{2273--2297}.
\PrintBackRefs{\CurrentBib}

\bibitem [\protect \citeauthoryear {%
S\BPBI J.~Smith%
\ \BBA {} Rasch%
}{%
S\BPBI J.~Smith%
\ \BBA {} Rasch%
}{%
{\protect \APACyear {2013}}%
}]{%
smith2013long}
\APACinsertmetastar {%
smith2013long}%
\begin{APACrefauthors}%
Smith, S\BPBI J.%
\BCBT {}\ \BBA {} Rasch, P\BPBI J.%
\end{APACrefauthors}%
\unskip\
\newblock
\APACrefYearMonthDay{2013}{}{}.
\newblock
{\BBOQ}\APACrefatitle {The long-term policy context for solar radiation
  management} {The long-term policy context for solar radiation
  management}.{\BBCQ}
\newblock
\APACjournalVolNumPages{Climatic Change}{121}{3}{487--497}.
\PrintBackRefs{\CurrentBib}

\bibitem [\protect \citeauthoryear {%
Stocker%
\ \protect \BOthers {.}}{%
Stocker%
\ \protect \BOthers {.}}{%
{\protect \APACyear {2013}}%
}]{%
stocker2013climate}
\APACinsertmetastar {%
stocker2013climate}%
\begin{APACrefauthors}%
Stocker, T\BPBI F.%
, Qin, D.%
, Plattner, G\BHBI K.%
, Tignor, M.%
, Allen, S\BPBI K.%
, Boschung, J.%
\BDBL {}others%
\end{APACrefauthors}%
\unskip\
\newblock
\APACrefYearMonthDay{2013}{}{}.
\newblock
\APACrefbtitle {Climate change 2013: The physical science basis.} {Climate
  change 2013: The physical science basis.}
\newblock
\APACaddressPublisher{}{Cambridge University Press Cambridge}.
\PrintBackRefs{\CurrentBib}

\bibitem [\protect \citeauthoryear {%
Tilmes%
, Sanderson%
\BCBL {}\ \BBA {} O'Neill%
}{%
Tilmes%
\ \protect \BOthers {.}}{%
{\protect \APACyear {2016}}%
}]{%
tilmes2016climate}
\APACinsertmetastar {%
tilmes2016climate}%
\begin{APACrefauthors}%
Tilmes, S.%
, Sanderson, B.%
\BCBL {}\ \BBA {} O'Neill, B.%
\end{APACrefauthors}%
\unskip\
\newblock
\APACrefYearMonthDay{2016}{}{}.
\newblock
{\BBOQ}\APACrefatitle {Climate impacts of geoengineering in a delayed
  mitigation scenario} {Climate impacts of geoengineering in a delayed
  mitigation scenario}.{\BBCQ}
\newblock
\APACjournalVolNumPages{Geophysical Research Letters}{43}{15}{8222--8229}.
\PrintBackRefs{\CurrentBib}

\bibitem [\protect \citeauthoryear {%
Williamson%
\ \BBA {} Turley%
}{%
Williamson%
\ \BBA {} Turley%
}{%
{\protect \APACyear {2012}}%
}]{%
williamson2012ocean}
\APACinsertmetastar {%
williamson2012ocean}%
\begin{APACrefauthors}%
Williamson, P.%
\BCBT {}\ \BBA {} Turley, C.%
\end{APACrefauthors}%
\unskip\
\newblock
\APACrefYearMonthDay{2012}{}{}.
\newblock
{\BBOQ}\APACrefatitle {Ocean acidification in a geoengineering context} {Ocean
  acidification in a geoengineering context}.{\BBCQ}
\newblock
\APACjournalVolNumPages{Philosophical Transactions of the Royal Society A:
  Mathematical, Physical and Engineering Sciences}{370}{1974}{4317--4342}.
\PrintBackRefs{\CurrentBib}

\bibitem [\protect \citeauthoryear {%
Zickfeld%
, MacDougall%
\BCBL {}\ \BBA {} Matthews%
}{%
Zickfeld%
\ \protect \BOthers {.}}{%
{\protect \APACyear {2016}}%
}]{%
zickfeld2016proportionality}
\APACinsertmetastar {%
zickfeld2016proportionality}%
\begin{APACrefauthors}%
Zickfeld, K.%
, MacDougall, A\BPBI H.%
\BCBL {}\ \BBA {} Matthews, H\BPBI D.%
\end{APACrefauthors}%
\unskip\
\newblock
\APACrefYearMonthDay{2016}{}{}.
\newblock
{\BBOQ}\APACrefatitle {On the proportionality between global temperature change
  and cumulative CO2 emissions during periods of net negative CO2 emissions}
  {On the proportionality between global temperature change and cumulative co2
  emissions during periods of net negative co2 emissions}.{\BBCQ}
\newblock
\APACjournalVolNumPages{Environmental Research Letters}{11}{5}{055006}.
\PrintBackRefs{\CurrentBib}

\end{thebibliography}

%
%
%
%
%

\end{document}